\RequirePackage{xcolor}
\documentclass[journal,twoside,web]{ieeecolor}
\usepackage{generic}
\usepackage{cite}
\usepackage{amsmath,amssymb,amsfonts}
\usepackage{algorithmic}
\usepackage{graphicx}
\usepackage{algorithm,algorithmic}
\usepackage{xurl}
\makeatletter
\let\NAT@parse\undefined
\makeatother
\usepackage[hidelinks]{hyperref}
\usepackage{textcomp}
\usepackage{booktabs,multirow,multicol}
\usepackage{subcaption}
\usepackage{etoolbox}
\usepackage{orcidlink}
\usepackage{epstopdf}

\def\BibTeX{{\rm B\kern-.05em{\sc i\kern-.025em b}\kern-.08em
    T\kern-.1667em\lower.7ex\hbox{E}\kern-.125emX}}
\providecommand{\refname}{References}

\makeatletter
\patchcmd{\thebibliography}
  {\section*{\color{black}}}
  {\section*{\color{black}\refname}}
  {}{}
\makeatother

\begin{document}

\title{End-to-End Markov State Sequence Learning for Auditory Attention Decoding}

\author{%
Yushan Yashengjiang\,\orcidlink{0009-0001-3497-7341},
Jie Zhang\,\orcidlink{0000-0003-1124-0854},
Miao Sun\,\orcidlink{0000-0001-8614-3704},
Huadong Liang,\\
\makebox[\textwidth][c]{Xin~Li, and Zhen-Hua~Ling\,\orcidlink{0000-0001-7853-5273},~\IEEEmembership{Senior~Member,~IEEE}}%
\thanks{Yushan Yashengjiang, Jie Zhang, and Zhen-Hua Ling are with the NERC-SLIP, University of Science and Technology of China (USTC), Hefei 230027, China (e-mail: yusan0426@mail.ustc.edu.cn; jzhang6@ustc.edu.cn; zhling@ustc.edu.cn). Corresponding author: Jie Zhang.}%
\thanks{Miao Sun is with the School of Information and Communication Engineering, Guangzhou Maritime University, Guangzhou 510725, China (e-mail: m.sun.1@hotmail.com).}%
\thanks{Huadong Liang is with the Artificial Intelligence Research Institute, iFLYTEK Company, Ltd., Hefei 230088, China (e-mail: hdliang@iflytek.com).}%
\thanks{Xin Li is with the Artificial Intelligence Research Institute, iFLYTEK Company, Ltd., Hefei 230088, China, and also with the School of Information Science and Technology, University of Science and Technology of China, Hefei 230052, China (e-mail: leexin@ustc.edu.cn).}}

\maketitle

\begin{abstract}
Auditory attention decoding (AAD) identifies the speaker a listener attends to from neural responses like electroencephalography (EEG), making it a key algorithm in neuro-steered hearing aids. However, most neural AAD models are trained as independent short-window classifiers, despite auditory attention being a temporally persistent cognitive state and short-window EEG--audio evidence often being noisy and ambiguous. We propose an end-to-end Markov AAD framework based on conditional random field (CRF) that trains window-level neural emissions under a two-state attention prior. The framework treats the logits of any AAD backbone as Markov emissions, learns the transition rate from a standard HMM initialization, and jointly optimizes cross-entropy and CRF objectives, allowing temporal continuity to guide representation learning rather than merely smoothing predictions after training. We also introduce ESCNet, an EEG--speech correlation backbone that preserves time-aligned features and converts the difference between two mean Pearson correlations into state logits. We evaluate the framework with four emission backbones spanning correlation-based, convolutional, recurrent, and attention-based designs. On the dynamic AVGC dataset, CRF training generally outperforms post-hoc HMM smoothing; with ESCNet, it achieves $86.5\%$ causal and $92.4\%$ non-causal accuracy using $1$s windows. On the static KUL and USTC datasets, it improves causal decoding over fixed-rate post-hoc HMM baselines by $5.6\%$ and $2.0\%$, respectively, showing the superiority of learning AAD as attention state sequence over isolated-window classification.

\end{abstract}

\begin{IEEEkeywords}
Auditory attention decoding, conditional random field, electroencephalography, hidden Markov model, neuro-steered hearing aids.
\end{IEEEkeywords}


\section{Introduction}

Auditory attention decoding (AAD) aims to identify the speaker that the listener attends to from neural responses recorded in  multi-talker scenarios \cite{MesgaraniChang2012,ZionGolumbic2013,OSullivan2015}. This is required by neuro-steered hearing aids (HAs): the decoded attention can guide gain control and thus help speech understanding of the target speaker, but an inaccurate or delayed decision may amplify noise sources \cite{GeirnaertTNSRE2020,Zhang2023BASEN,Xu2026USTC}. Electroencephalography (EEG) is particularly attractive for AAD because its high time resolution provides continuous evidence of selective attention. Recent studies on EEG-based AAD include end-to-end EEG--audio decoding, short-decision windows, speech--EEG match--mismatch learning, and multi-direction spatial decoding \cite{Nguyen2024AADNet,Shi2025TNSRE,Zhang2024NERCSLIP,Zhang2025TFMatchMismatch,Zhang2026NeuroCapsNet}. A practical AAD algorithm, however, must satisfy two requirements: resisting noisy short-term neural evidence and remaining responsive to genuine attention switches \cite{Alickovic2019,Zhu2024EarEEG,RealTimeBCSHNature2026}.

Most existing AAD problems were formulated as independent classifiers over short analysis windows. Given an EEG segment and one or more candidate speech representations, they produce a window-level compatibility score and select the most likely attended stream \cite{Mirkovic2015,deCheveigne2018CCA,Ciccarelli2019,Xu2026USTC}, e.g., see stimulus-reconstruction and correlation-based pipelines \cite{OSullivan2015,Crosse2016,Das2016KUL} as well as recent convolutional, recurrent and attention-based neural decoders \cite{Vandecappelle2021,Accou2021,Kuruvila2021,Borsdorf2024LSTM,Nguyen2024AADNet}. Although these methods differ substantially in how they extract EEG--audio correspondences, their dominant training objective treats neighboring windows as separate samples and therefore omits the temporal dynamics of auditory attention.
This omission causes a structural mismatch between model training and the underlying decoding task. Auditory attention usually persists over extended periods and changes only at sparse moments, whereas its short-window neural signature is noisy and can be locally ambiguous \cite{ShinnCunningham2008,Rimmele2015,Akram2016,Geirnaert2021}. Under independent-window supervision, an isolated erroneous decision and a sustained trajectory error incur the same local penalty. The learned scores can thus be discriminative for individual windows without providing well-formed evidence for tracking a persistent cognitive state, particularly at short decision windows or near attention switches \cite{Rotaru2024AVGC,Heintz2025AADHMM}.

Temporal post-processing can partially address this issue but leaves the training mismatch unsolved. Hidden Markov models (HMMs) support both causal and non-causal inference, combining window-level scores with a transition prior to suppress implausible rapid changes in the decoded state \cite{VanEyndhoven2017,Geirnaert2021,Heintz2025AADHMM}. In the conventional post-hoc pipeline, however, the EEG--audio encoder is optimized before the state model is introduced. The sequence model can only smooth the supplied emissions; it cannot shape the representation from which those emissions are obtained. Consequently, temporal continuity improves inference without teaching the encoder what constitutes reliable evidence for an attention trajectory.

In this work, we therefore formulate AAD as end-to-end Markov state learning. We interpret the output logits of any window-level AAD backbone as neural emissions of a two-state conditional random field (CRF) and jointly optimize them using a combined cross entropy (CE) and CRF objective \cite{SuttonMcCallum2012,MaHovy2016,Mehta2022NeuralHMM}. By contrasting the ground-truth state trajectory with all competing trajectories under an explicit transition prior, CRF training propagates sequence-level supervision into the emission backbone. The resulting single trained model can support both causal inference for online AAD and non-causal inference for offline analysis without retraining. To provide time-resolved EEG--speech emissions, we propose the ESCNet model, which preserves the temporal axis in both encoders and maps the difference between two mean correlation scores directly to the two attention-state logits.
We evaluate the proposed framework on the dynamic AVGC dataset and the static KUL and USTC datasets given multiple decision-window lengths and inference modes. To test if the formulation is tied to one emission architecture, we compare four heterogeneous backbones: ESCNet, AADNet~\cite{Nguyen2024AADNet}, LSTM~\cite{Borsdorf2024LSTM}, and Attn-GRU~\cite{Accou2021,Kuruvila2021}. Comparisons with post-hoc HMM baselines under common backbones, data splits, and preprocessing show that incorporating the Markov structure during training can improve the learned emissions beyond merely smoothing their outputs at inference time. A fixed-transition control further isolates this sequence-training effect from learning the transition rate. These results support the central premise that AAD should be learned as attention-state sequence estimation rather than as isolated-window classification. 

The contribution of this work is fourfold: 1) We recast AAD as attention-state sequence estimation and identify the mismatch between independent-window training and temporally persistent attention; 2) We propose an end-to-end CRF framework that propagates transition-aware sequence supervision into neural EEG--audio emissions while supporting both causal and non-causal inference; 3) We introduce ESCNet, a time-preserving EEG--speech correlation backbone that produces interpretable two-state emissions without extra classification heads; and 4) Controlled comparisons with post-hoc HMM inference on multiple datasets reveal the superiority of the proposed framework. The reproducible source code of this work is publicly available at
{\color{blue}\url{https://github.com/YusanX/AAD-CRF}}.

\section{Related Work}

\textbf{EEG-based AAD:}
Research on selective listening showed that cortical responses track attended speech more strongly than competing speech in multi-talker cases~\cite{MesgaraniChang2012,ZionGolumbic2013,DingSimon2012}. This motivates stimulus-reconstruction, encoding and correlation-based methods that estimate an attended-speech representation (e.g., speech envelope) from EEG and identify the candidate stream with the strongest neural correspondence \cite{OSullivan2015,Crosse2016,Fuglsang2017,Das2016KUL,Xu2026USTC}. More recent convolutional, recurrent, attention-based and multimodal networks learn this correspondence directly from data \cite{Ciccarelli2019,Vandecappelle2021,Accou2021,Kuruvila2021,Borsdorf2024LSTM,Nguyen2024AADNet}. Related developments include speech--EEG match--mismatch learning, multi-direction spatial attention decoding, self-supervised speech representations, and compact EEG configurations \cite{Zhang2024NERCSLIP,Zhang2025TFMatchMismatch,Zhang2026NeuroCapsNet,Han2023SSL,Zhu2024EarEEG}. Despite their different representations and architectures, these approaches predominantly optimize a label or compatibility score for each analysis window independently. Our work complements this line with two coupled contributions: ESCNet provides time-preserving, correlation-based window emissions, while CRF training uses the temporal structure of the attention trajectory to shape those emissions.

\textbf{Brain-assisted speech enhancement (BASE):}
BASE uses neural attention clues to extract the attended speaker from an acoustic mixture. For example, a typical BASE model performs time-domain enhancement through convolutional cross-attention between EEG and acoustic features~\cite{Zhang2023BASEN}. Subsequent analysis decomposes end-to-end BASE into implicit AAD, speech separation, and target selection, identifying the AAD mechanism as an important performance bottleneck \cite{Xu2026USTC}. Other studies reduce the EEG sensing complexity using end-to-end, geometry-constrained, or subject-adaptive channel selection tricks~\cite{Xu2024Gumbel,Zuo2025Geometry,Xu2025SubjectAdaptive}. These methods establish the downstream importance of reliable neural attention evidence, but their primary objectives are speech enhancement or channel selection. In contrast, in this work we focus on the temporal structure of the decoded attention itself and train the AAD emissions under an explicit state-transition model.

\textbf{Temporal modeling for AAD:}
State-space methods exploit the observation that auditory attention is persistent and switches relatively infrequently. Both causal and non-causal HMM inference combine noisy window-level evidence with a transition prior, thereby reducing implausible state fluctuations and improving attention tracking performance~\cite{Rabiner1989,VanEyndhoven2017,Geirnaert2021,Heintz2025AADHMM,RealTimeBCSHNature2026}. Existing AAD pipelines generally introduce this temporal model after training the neural decoder: the encoder is learned from independent windows, and the state model subsequently smooths its outputs. This separation prevents temporal inference from influencing how the EEG--audio evidence is represented. Our framework retains the same interpretable form of Markov prior but incorporates it into training, allowing trajectory-level supervision to update the emission encoder directly.

\textbf{Sequence-discriminative learning:}
CRFs and related sequence-discriminative criteria couple local evidence with structured decisions by scoring the reference path against competing paths \cite{Lafferty2001,Povey2008,SuttonMcCallum2012}. Neural extensions, including LSTM--CRF and neural-HMM models, propagate this structured supervision into learned representations and have been widely used when independent frame or token objectives are misaligned with sequence-level inference \cite{MaHovy2016,Mehta2022NeuralHMM}. We introduce this principle into AAD using a two-state CRF formulation and evaluate it against post-hoc HMM inference under common backbones and protocols. A fixed-transition control isolates the central distinction of our work: whether attention-state dynamics are used only to refine predictions or also to supervise the neural evidence from which those predictions are made.

\begin{figure*}[t]
\centering
\includegraphics[width=\textwidth]{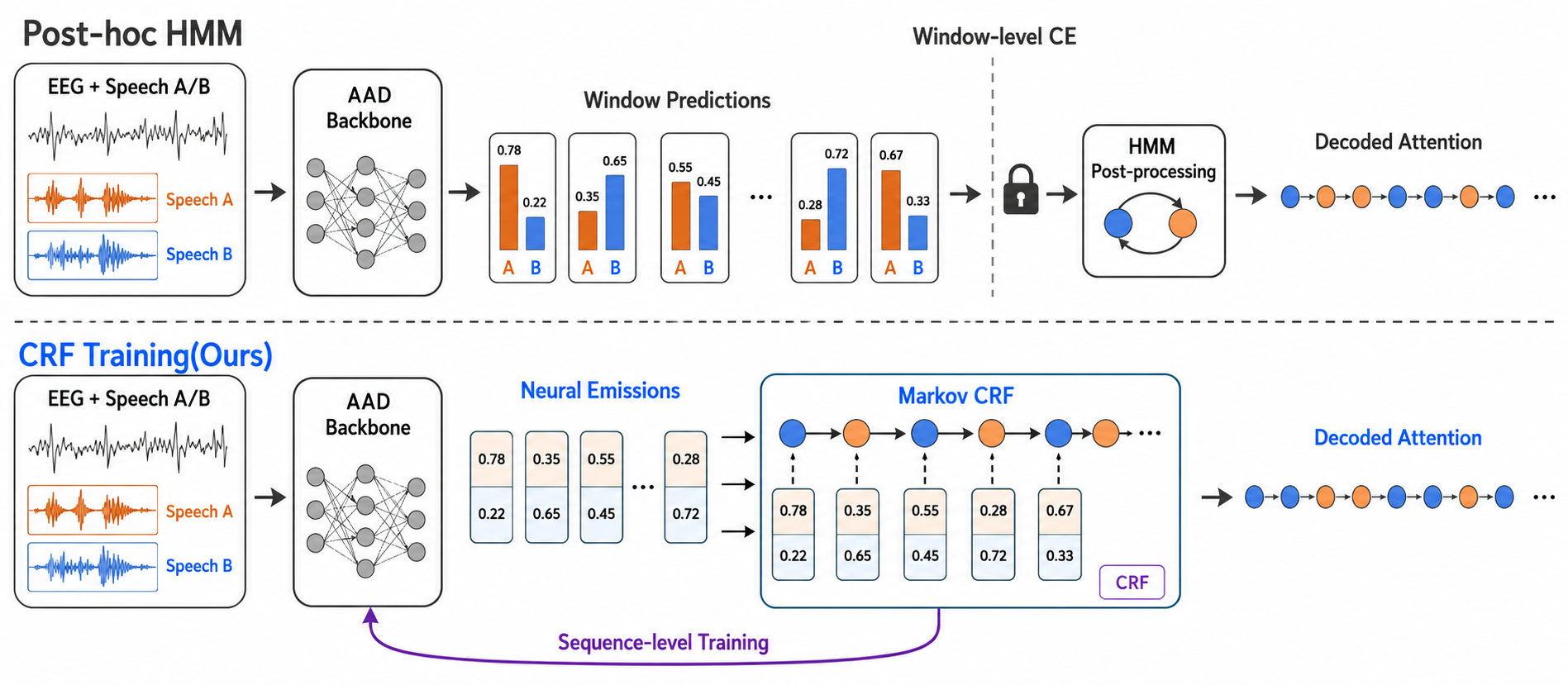}
\caption{Comparison between conventional post-hoc HMM inference and the proposed CRF framework. In the post-hoc pipeline (top), the AAD backbone is trained only with window-level CE and remains fixed when the HMM is applied. In our framework (bottom), the backbone outputs neural emissions for a Markov CRF, and the sequence-level objective backpropagates through the CRF into the backbone. The resulting emissions support both causal and non-causal decoding.}
\label{fig:framework}
\end{figure*}

\section{Methodology}
\label{sec:method}

The overall framework is shown in Fig.~\ref{fig:framework}. The upper part shows conventional post-hoc HMM inference: an AAD backbone is first optimized using the CE loss, then frozen, and its independent predictions are smoothed by an HMM. The lower shows the proposed CRF training method, where the backbone logits become neural emissions of a Markov CRF and the sequence objective directly supervises the backbone.

The proposed framework separates local evidence extraction from temporal state modeling while training them jointly. A window-level backbone, instantiated primarily by ESCNet, first produces two state logits. The CRF combines these emissions with an explicit transition prior, compares the ground-truth attention trajectory against competing trajectories, and propagates the resulting sequence-level gradient into the backbone. The following subsections define the emission interface, ESCNet, the Markov state model, CRF training, and inference.

\subsection{Problem Formulation}

We formulate AAD as estimating a latent attention-state sequence rather than a set of independent window labels. A listening trial is sliced into $N$ consecutive windows of length $T$ samples with a fixed stride; for window $n$ we observe an EEG segment $X_n\in\mathbb{R}^{C\times T}$ over $C$ channels and two candidate speech envelopes $a_n^{(0)},a_n^{(1)}\in\mathbb{R}^{T}$, one per competing talker. The attended talker at window $n$ is a hidden state $z_n\in\{0,1\}$, and the quantity we ultimately decode is the whole trajectory $z_{1:N}$. Writing the observations as $o_n=(X_n,a_n^{(0)},a_n^{(1)})$, our formulation factorizes the trajectory posterior $p(z_{1:N}\mid o_{1:N})$ into a window-level \emph{emission} term and a \emph{transition} term that encodes the temporal persistence of attention \cite{Rabiner1989,Heintz2025AADHMM}.

The proposed CRF framework supports different window-level AAD backbones via a common two-logit interface. A multimodal backbone may use $(X_n,a_n^{(0)},a_n^{(1)})$, and an EEG-only backbone may ignore the candidate-envelope inputs. The CRF couples the resulting logits through a two-state Markov attention model and propagates trajectory-level supervision back into the backbone, so the learned scores support sequence decoding rather than only isolated-window classification.

\subsection{Neural Emission Interface}

The CRF accepts any window-level AAD backbone that produces two logits for the competing attention states. We therefore define a common interface that converts the backbone outputs into the log-emission scores used by the Markov model. Without loss of generality, we consider the multimodal backbone $h_\psi$ as an illustrative example, which outputs logits
\begin{equation}
e_n=h_\psi(X_n,a_n^{(0)},a_n^{(1)})\in\mathbb{R}^{2}.
\label{eq:generic-emission}
\end{equation}
We convert these logits into log-emission scores as
\begin{equation}
b_n=\log\operatorname{softmax}(e_n),
\label{eq:log-emission}
\end{equation}
where $b_n(j)$ is the neural evidence assigned to state $z_n=j$. This interface is deliberately minimal: the backbone may be a correlation-based AAD network, a convolutional AAD model, a recurrent decoder, or an EEG-only classifier, provided that it exposes window-level logits over multiple attention states.

\subsection{ESCNet: EEG--Speech Correlation Network}
\label{sec:escnet}

\begin{figure}[t]
\centering
\includegraphics[width=\linewidth]{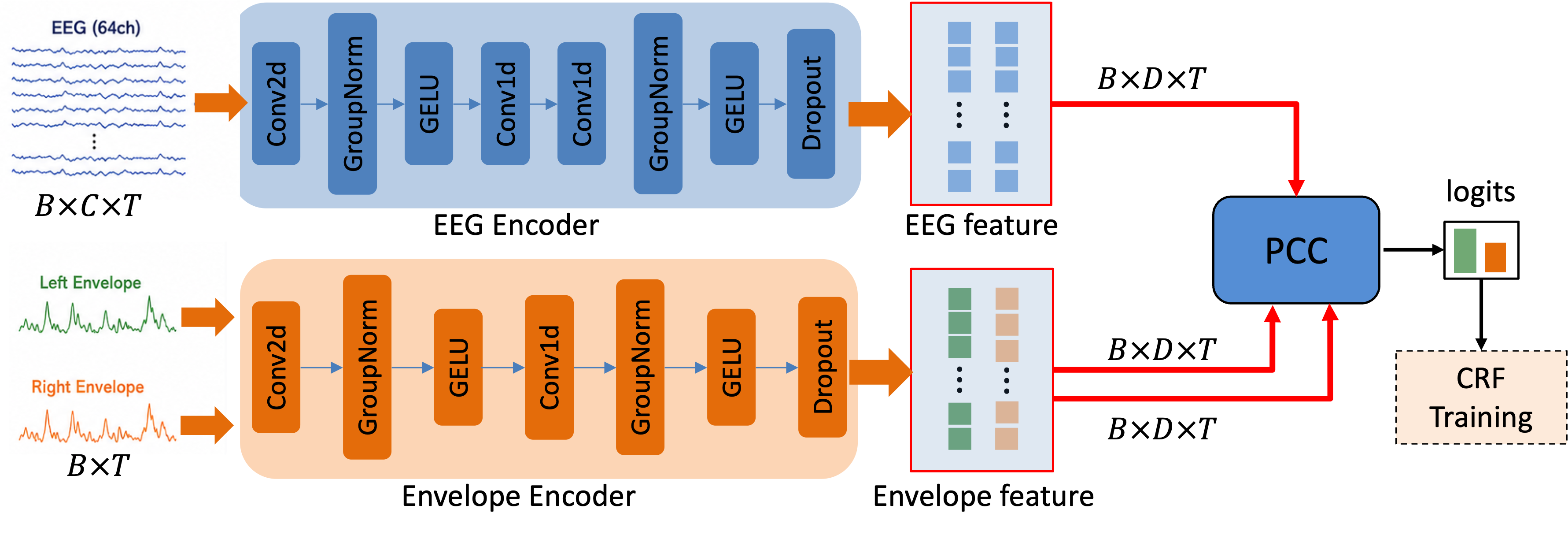}
\caption{Proposed ESCNet AAD backbone, where EEG signals and two candidate speech envelopes are encoded as time-aligned feature sequences, and their mean temporal PCCs are compared to produce two-state logits.}
\label{fig:escnet}
\end{figure}

Short-window AAD requires an emission backbone that retains the temporal correspondence between cortical responses and speech envelopes. We therefore propose ESCNet, a lightweight EEG--speech correlation network that preserves the temporal axis throughout feature extraction and converts neural tracking differences directly into attention-state logits. As illustrated in Fig.~\ref{fig:escnet}, ESCNet comprises an EEG encoder, a shared speech-envelope encoder, and a parameter-free Pearson correlation coefficient (PCC) scoring stage.

Given an EEG window $X_n$, the EEG encoder $f_\theta$ first applies a two-dimensional spatial convolution across channels and then uses one-dimensional temporal convolutions to extract neural tracking features. Group normalization and GELU activation follow the convolutional stages, and dropout regularizes the output. A speech-envelope encoder $g_\phi$ applies a corresponding convolutional stack to each candidate envelope; its weights are shared across candidates so that their scores remain directly comparable. The two encoders preserve the window timeline and produce $U_n=f_\theta(X_n)\in\mathbb{R}^{D\times T}$ and $V_n^{(k)}=g_\phi(a_n^{(k)})\in\mathbb{R}^{D\times T}$ for $k\in\{0,1\}$.

ESCNet computes the mean temporal PCC between each EEG--speech feature pair as
\begin{equation}
\operatorname{sim}(U,V)=\frac{1}{D}\sum_{d=1}^{D}
\frac{\langle \bar U_{d},\,\bar V_{d}\rangle}{\lVert \bar U_{d}\rVert\,\lVert \bar V_{d}\rVert},
\label{eq:track}
\end{equation}
where $\bar U_d,\bar V_d$ are the time-centered $d$-th feature rows, respectively. Then, we convert the difference between the two candidate scores into antisymmetric state logits, e.g.,
\begin{align}
\delta_n &=\operatorname{sim}(U_n,V_n^{(0)})-\operatorname{sim}(U_n,V_n^{(1)}),\\
\ell_n&=[\,\delta_n,\,-\delta_n\,].
\label{eq:logit}
\end{align}
Unlike AADNet, which feeds a vector of channel-wise correlations to a learned classifier, ESCNet uses the signed difference between the two mean correlations as its decision variable. The antisymmetric logits give the two attention states a common score scale without an additional classification head. Setting $e_n=\ell_n$ and applying Eq.~\eqref{eq:log-emission} yields the normalized log-emission scores used by the CRF; the subsequent Markov model, training objective, and inference algorithms remain unchanged for the other evaluated backbones.

\subsection{Markov Attention-State Model}

A purely window-level emission ignores the fact that auditory attention is a slowly varying state, which persists for many windows and changes only at sparse moments \cite{ShinnCunningham2017,Geirnaert2021}. Let $q$ denote a base switching rate per second and let $\tau$ denote the decision-window duration in seconds. Following HMM-based AAD post-processing \cite{Heintz2025AADHMM}, we convert the rate into the per-window switching probability $q_{\tau}=q\tau$. The resulting two-state transition matrix is
\begin{equation}
A_{\tau}=\begin{bmatrix}1-q_{\tau} & q_{\tau}\\[2pt] q_{\tau} & 1-q_{\tau}\end{bmatrix},
\qquad q_{\tau}=q\tau\in(0,\tfrac12),
\label{eq:trans}
\end{equation}
together with a uniform initial distribution $\pi=[\tfrac12,\tfrac12]$. Scaling by $\tau$ keeps the switching prior on a common physical-time scale across the $1$, $2$, and $4$\,s settings: longer windows span proportionally more time and therefore receive a proportionally larger transition probability. A small $q_{\tau}$ still encodes the prior that the attended talker rarely changes between adjacent decisions, suppressing isolated prediction flips that dominate window-independent decoding.

For HMM-post, we fix the base rate at $q_0=10^{-3}$ and use $q_{\tau}=q_0\tau$ for every window-length setting, similar as \cite{Heintz2025AADHMM}. For CRF, the base rate is initialized from the same value, $q=q_0$, and is then learned jointly with the neural emissions during sequence training; the learned rate is likewise converted to $q_{\tau}=q\tau$ before constructing $A_{\tau}$. The corresponding fixed or learned transition is reused for causal and non-causal decoding at inference. Thus, the main comparison evaluates the conventional fixed-prior HMM-post pipeline against the complete end-to-end CRF framework, while the fixed-$q$ CRF control in Sec.~\ref{sec:exp-results} isolates the contribution of sequence-aware emission learning from transition-rate adaptation.

\subsection{End-to-End CRF Training}
\label{sec:crf-training}

As shown by the lower pathway of Fig.~\ref{fig:framework}, CRF training connects the sequence objective directly to the emission backbone. Training a backbone on independent per-window labels is misaligned with trajectory decoding, because such a loss never tells the encoder which emissions lead to a coherent state sequence \cite{Lafferty2001,MaHovy2016}. We instead optimize the CRF objective defined over the whole trajectory, which scores the ground-truth path against all competing paths under the current emissions and transition. For a label trajectory $z_{1:N}$, the un-normalized log-score is given by
\begin{equation}
S(z_{1:N})=\log\pi_{z_1}+\sum_{n=1}^{N} b_n(z_n)+\sum_{n=2}^{N}\log (A_{\tau})_{z_{n-1},z_n},
\label{eq:score}
\end{equation}
and the sequence-discriminative loss for a trial with ground truth $z_{1:N}^{\ast}$ is the length-normalized negative conditional log-likelihood
\begin{equation}
\mathcal{L}_{\mathrm{CRF}}
=-\frac{1}{N}\Big(S(z_{1:N}^{\ast})-\operatorname*{logsumexp}_{z_{1:N}} S(z_{1:N})\Big),
\label{eq:crf}
\end{equation}
where the partition term is evaluated in $O(N)$ by the forward recurrence and is fully differentiable. The gradient of Eq.~\eqref{eq:crf} drives the model toward $(\text{empirical marginals along }z^{\ast})-(\text{model posterior marginals})$, so it raises the emission scores that support the correct trajectory while lowering those that support competing trajectories under the same transition prior \cite{Povey2008,Mehta2022NeuralHMM}. Since $b_n$ is computed from the backbone logits in Eq.~\eqref{eq:log-emission}, this trajectory-level signal back-propagates into the emission network, helping the backbone learn evidence that is coherent under the Markov state model rather than merely locally discriminative.

We train in two phases to initialize locally discriminative emissions before applying sequence supervision. During warm-up, we minimize only the per-window CE loss $\mathcal{L}_{\mathrm{CE}}$ with respect to the window labels. We then jointly optimize the local and trajectory-level objectives using
\begin{equation}
\mathcal{L}=\lambda_{\mathrm{CRF}}\,\mathcal{L}_{\mathrm{CRF}}+\lambda_{\mathrm{CE}}\,\mathcal{L}_{\mathrm{CE}},
\label{eq:total}
\end{equation}
where the retained CE term anchors per-window calibration while the CRF term enforces trajectory coherence. During this joint phase, the base switching rate $q$ is optimized together with the emission backbone after being initialized at $q_0=10^{-3}$. The CRF term is the supervised sequence-discriminative criterion used in LSTM--CRF and neural-HMM models, here specialized to an interpretable two-state attention chain \cite{Lafferty2001,MaHovy2016,Mehta2022NeuralHMM}. It differs from a marginal-likelihood (Baum--Welch) objective~\cite{Rabiner1989}, which would maximize $\log p(o_{1:N})$ without reference to the labels and leave label supervision to the CE term alone. As indicated by the feedback arrow in Fig.~\ref{fig:framework}, gradients from $\mathcal{L}_{\mathrm{CRF}}$ pass through the neural emissions and update both the backbone and the transition rate jointly with the local CE term.

\subsection{Inference}
At the model inference stage, we decode the attention trajectory using the transition associated with each training path: HMM-post uses the fixed base rate $q_0$, whereas CRF uses the rate learned from the same initialization during sequence training. In both cases, the base rate is scaled by the test window duration through $q_{\tau}=q\tau$. In the causal mode, we run the forward filter, which produces the posterior $p(z_n\mid o_{1:n})$ from past and current windows only and is suitable for real-time neuro-steered HA systems \cite{Heintz2025AADHMM,RealTimeBCSHNature2026}. In the non-causal mode, the entire held-out trial is available: forward messages $\alpha_n\!\propto\! p(z_n,o_{1:n})$ are combined with backward messages $\beta_n\!\propto\! p(o_{n+1:N}\mid z_n)$ to form the full-sequence posterior $p(z_n\mid o_{1:N})\propto\alpha_n\beta_n$ \cite{Rabiner1989}. In both modes, the decoded state is $\hat z_n=\arg\max_{j} p(z_n=j\mid\cdot)$. Reporting both modes separates causal online decoding from the additional benefit of non-causal access to future windows, and the same CRF-trained model serves both settings without retraining.

\section{Experimental Setup}\label{sec:exp-setup}
In this section, we present the experimental setup including datasets, backbones, metrics and implementation details.

\subsection{Datasets}
In order to see whether end-to-end Markov training improves AAD by learning better emissions rather than merely smoothing window-level outputs after training, we evaluate the same CRF sequence objective in three settings: dynamic attention switching on the AVGC dataset~\cite{Rotaru2024AVGC}, static decoding on the KUL~\cite{Das2016KUL} and USTC~\cite{Xu2026USTC} datasets, and transfer across multiple EEG--audio backbones. This directly shows the central claim that sequence supervision improves the emission model itself and should therefore remain useful across datasets, window lengths, and architectures.

AVGC is taken as the main dataset because it includes trial-internal attention switches, whereas KUL and USTC serve as static AAD datasets. After consent-based exclusions, the public AVGC release contains $13$ normal-hearing participants, each with six $10$-min trials, $64$-channel EEG, and two competing speech envelopes sampled at $128$\,Hz \cite{Rotaru2024AVGC}. Each trial includes one programmed spatial attention switch at approximately $5$\,min; we obtain its time and the attended stream before and after the switch from the trial-level annotations.

AVGC is evaluated using participant-dependent leave-one-trial-out (LOTO) cross-validation. For each participant, one of the six trials is held out for testing, a new model is initialized from scratch and trained on that participant's other five trials, and this procedure is repeated for all six choices of test trial. The protocol therefore comprises $13\times6=78$ independently trained fold models and evaluates participant-dependent decoding rather than cross-participant generalization. No separate validation split or early stopping is used; every fold follows the fixed $50$-epoch schedule specified below.

KUL and USTC contain static AAD trials and can therefore test if Markov training remains useful when the dominant temporal structure is state persistence rather than explicit switching \cite{Das2016KUL,Xu2026USTC}. KUL contains $16$ normal-hearing participants, each completing eight $6$-min trials with two competing stories and $64$-channel EEG. USTC is a dichotic-listening corpus in which each of $18$ participants attended to one of two simultaneous Mandarin news speakers across $20$ trials of $120$\,s. It provides $64$-channel EEG recorded at $8196$\,Hz, which we downsample to $128$\,Hz. After removing four USTC trials with duplicated speech pairs, we evaluate the remaining $356$ trials using participant-dependent LOTO folds, matching the KUL protocol \cite{Xu2026USTC}. Within each dataset, all methods use identical preprocessing, folds, window labels, and evaluation windows; performance differences therefore reflect the training and inference strategies rather than protocol changes.
\subsection{Backbones and Metrics}

Comparison AAD models are chosen to distinguish framework-level gains from architecture-specific gains. We evaluate four neural emission networks: temporal-correlation, convolutional, recurrent, and attention-based designs.

\textbf{ESCNet} is the correlation-based backbone proposed in Sec.~\ref{sec:escnet}. It serves as the primary instantiation for controlled and statistical analyses, while other backbones test whether CRF training transfers beyond the proposed emission structure.

\textbf{AADNet} \cite{Nguyen2024AADNet} is a convolutional model that encodes EEG and speech envelopes with parallel multi-scale convolutional branches. It computes channel-wise PCCs between the encoded EEG and speech features, then maps the resulting correlation vector to attention probabilities through a learned fully-connected classifier.

\textbf{LSTM} \cite{Borsdorf2024LSTM} is a recurrent network that encodes EEG with a joint spatiotemporal two-dimensional convolution block followed by time-distributed dense projections. It encodes each speech envelope with a shared one-dimensional convolution--LSTM branch and scores EEG--speech compatibility by averaging per-timestep cosine similarities over the window.

\textbf{Attn-GRU} \cite{Accou2021,Kuruvila2021} is an attention-based recurrent network adapted from the ICASSP 2023 Auditory EEG Challenge. It encodes EEG with a pre-normalized multi-head self-attention block followed by dilated convolutions and encodes speech envelopes with a bidirectional GRU followed by dilated convolutions. A learned linear head applied to the temporal cross-correlation matrix of the two feature sequences produces the EEG--speech compatibility score.

The primary accuracy metric is full-trial window accuracy under each inference mode. We first compute the fraction of correctly decoded windows in each held-out trial, average these fold accuracies within each participant, and then average across participants. For KUL and USTC, where trial-internal switches are not part of the standard protocol, we additionally report $95\%$ confidence intervals and paired participant-level improvements over HMM-post.

For AVGC, we also measure how quickly a decoded sequence responds to its annotated attention switch. Let $t_{\mathrm{sw}}$ be the annotated switch time, $t_n$ the start time of window $n$, and the switch-window index  $n^{\ast}=\min\{n:t_n\geq t_{\mathrm{sw}}\}$, respectively. For a specified output sequence, let $z^{+}$ denote the attended-talker state after the switch and  $\hat z_n$ the decoded state. We scan forward from $n^{\ast}$ and set $\hat n$ to the first index satisfying $\hat z_{\hat n}=z^{+}$. Because the decoding stride equals the window duration $\tau$, the switch delay is
\begin{equation}
D_{\mathrm{sw}}=(\hat n-n^{\ast})\tau .
\label{eq:switch-delay}
\end{equation}

\subsection{Implementation Details}
The AVGC dataset is evaluated with $1$, $2$, and $4$\,s decision windows to characterize the accuracy--responsiveness trade-off in the case of dynamic attention. The $1$\,s setting provides temporally precise but weak emissions, the $2$\,s setting balances evidence strength and temporal resolution, and the $4$\,s setting provides stronger evidence at the cost of blurring attention switches. KUL and USTC are reported at $1$\,s, where weak per-window evidence provides the most stringent test of whether sequence training improves static decoding. In all comparisons, the decoding stride matches the window length; after resampling to $128$\,Hz, the $1$, $2$, and $4$\,s AVGC settings contain $128$, $256$, and $512$ samples, respectively.
\begin{table*}[t]
\caption{Dynamic AAD results on the AVGC dataset. Causal decoding uses past and current windows, and non-causal (NC) decoding uses the complete trial. Switch delay (s) is the mean CRF-Causal delay from Eq.~\eqref{eq:switch-delay}, including undetected switches (smaller is more responsive). \textbf{Bold}: best CRF-NC result per window group. }
\label{tab:avgc-main}
\centering
\begin{tabular}{llccccccc}
\toprule
Window & Model & Raw$\uparrow$ & Post-Causal$\uparrow$ & Post-NC$\uparrow$ & CRF-Raw$\uparrow$ & CRF-Causal$\uparrow$ & CRF-NC$\uparrow$ & Sw.Delay$\downarrow$ \\
\midrule
\multirow{4}{*}{1 s} & ESCNet & 55.7 & 77.5 & 84.3 & 56.4 & 86.5 & \textbf{92.4} & 23.3 \\
    & AADNet    & 53.7 & 71.1 & 76.9 & 53.0 & 77.0 & 86.2 & 52.6 \\
    & LSTM      & 53.1 & 69.7 & 76.4 & 54.8 & 83.1 & 90.0 & 57.3 \\
    & Attn-GRU  & 53.9 & 58.9 & 62.1 & 54.7 & 73.3 & 79.9 & 15.2 \\
\midrule
\multirow{4}{*}{2 s} & ESCNet & 58.0 & 81.4 & 88.8 & 59.2 & 89.6 & \textbf{94.5} & 39.8 \\
    & AADNet    & 57.8 & 81.7 & 89.8 & 55.2 & 83.4 & 87.4 & 84.6 \\
    & LSTM      & 55.7 & 79.5 & 86.8 & 57.1 & 88.1 & 90.6 & 105.7 \\
    & Attn-GRU  & 57.4 & 63.9 & 68.1 & 57.6 & 80.2 & 87.1 & 22.4 \\
\midrule
\multirow{4}{*}{4 s} & ESCNet & 60.2 & 85.2 & 91.5 & 61.9 & 89.8 & \textbf{95.9} & 79.6 \\
    & AADNet    & 61.4 & 86.4 & 93.3 & 58.5 & 84.9 & 84.6 & 131.7 \\
    & LSTM      & 59.6 & 84.2 & 91.3 & 59.9 & 87.5 & 86.8 & 159.5 \\
    & Attn-GRU  & 61.4 & 69.6 & 75.0 & 61.9 & 84.5 & 90.1 & 33.7 \\
\bottomrule
\end{tabular}
\end{table*}

Each trial, rather than an encoder mini-batch, is the sequence unit for both training and decoding. During CE warm-up and joint CRF training, the training loop preserves the chronological order of all $N$ windows in a trial, assembles their logits into one $N\times2$ tensor, evaluates the loss once on that complete trial, and performs one optimizer update. To limit GPU memory, ESCNet and AADNet compute encoder outputs in contiguous chunks of at most $32$ windows; the chunk logits are differentiably concatenated before either the CE or CRF loss is evaluated, so gradients reach every window in the trial.  LSTM and Attn-GRU use their implementations' full-sequence setting and process the complete trial without encoder chunking.

All comparison methods use the same signal pre-processing and training schedule within a dataset. EEG is re-referenced to the common average, band-pass filtered from $1$ to $9$\,Hz with a fourth-order filter, and resampled to $128$\,Hz before windowing. End-to-end models are trained for $50$ epochs with AdamW, weight decay $10^{-4}$, gradient clipping at $1.0$, and a cosine learning-rate schedule. The first $10$ epochs use only the local CE warm-up; the remaining epochs optimize the combined loss in Eq.~\eqref{eq:total} with $\lambda_{\mathrm{CRF}}=5.0$ and $\lambda_{\mathrm{CE}}=0.5$. For HMM-post, the base switching rate is fixed at $q_0=10^{-3}$. For CRF, $q$ is initialized at $q_0$ and then learned jointly during the CRF phase. For every method and window length $\tau$, the transition matrix uses the scaled per-window probability $q_{\tau}=q\tau$, following \cite{Heintz2025AADHMM}. Learning rates are backbone-specific and fixed across window lengths: $3\times10^{-3}$ for ESCNet, $10^{-4}$ for AADNet, $3\times10^{-4}$ for LSTM, and $2\times10^{-4}$ for Attn-GRU. All experiments use a fixed random seed of $42$ and run on an NVIDIA GeForce RTX 4090 GPU.

We compare three training and inference paths for each backbone. The \emph{Raw} path trains the backbone with its standard window-level objective and applies argmax to the emission logits. The \emph{HMM-post} path applies causal or non-causal HMM inference with fixed $q_0$ to the same independently trained backbone, matching the standard post-processing paradigm \cite{Heintz2025AADHMM}. The \emph{CRF} path first warms up the backbone with its local objective and then jointly optimizes the sequence loss and the transition rate from Sec.~\ref{sec:crf-training}. From each model, we report the performance of raw emissions and causal/non-causal decoding.

\section{Performance Evaluation}\label{sec:exp-results}

\subsection{Dynamic AAD on AVGC}
Evaluation on the AVGC dataset tests the setting in which Markov state learning should matter most. Because AVGC contains explicit attention switches, the model must suppress isolated false flips while still responding when the attended speaker genuinely changes. Table~\ref{tab:avgc-main} is thus organized by window length and reports raw emission accuracy, fixed-rate post-hoc HMM inference, learned-rate CRF inference, and switch delay for each backbone. It evaluates the complete end-to-end framework against the conventional post-processing pipeline; the fixed-rate CRF control in Table~\ref{tab:ablations} then isolates the effect of sequence-aware emission learning.

Table~\ref{tab:avgc-main} reveals that the advantage of end-to-end sequence training is most consistent when short windows make local evidence most ambiguous. At $1$\,s, CRF-Causal exceeds Post-Causal for all four backbones by $5.9\%$--$14.4\%$, while CRF-NC exceeds Post-NC by $8.1\%$--$17.8\%$. The gains become more architecture-dependent as the window grows: ESCNet remains consistently better than its post-hoc counterpart in both inference modes at $1$, $2$, and $4$\,s (causal gains of $9.0\%$, $8.2\%$, and $4.6\%$; non-causal gains of $8.1\%$, $5.7\%$, and $4.4\%$), whereas AADNet and LSTM lose their non-causal advantage at $4$\,s. This contrast indicates that sequence training is most beneficial when the emission backbone produces temporally compatible evidence; with longer-window AADNet and LSTM emissions, the very large switch delays suggest over-persistence that can offset the gain from non-causal decoding. ESCNet consequently achieves the best CRF-NC accuracy in every window group ($92.4\%$, $94.5\%$, and $95.9\%$), even though its CRF-Raw accuracy is only $56.4\%$--$61.9\%$. The $34.0\%$--$36.0\%$ improvement from raw to non-causal decoding underscores that isolated-window decisions remain difficult and that trajectory-level decoding can resolve much of this local ambiguity.

\begin{figure}[t]
\centering
\includegraphics[width=\linewidth]{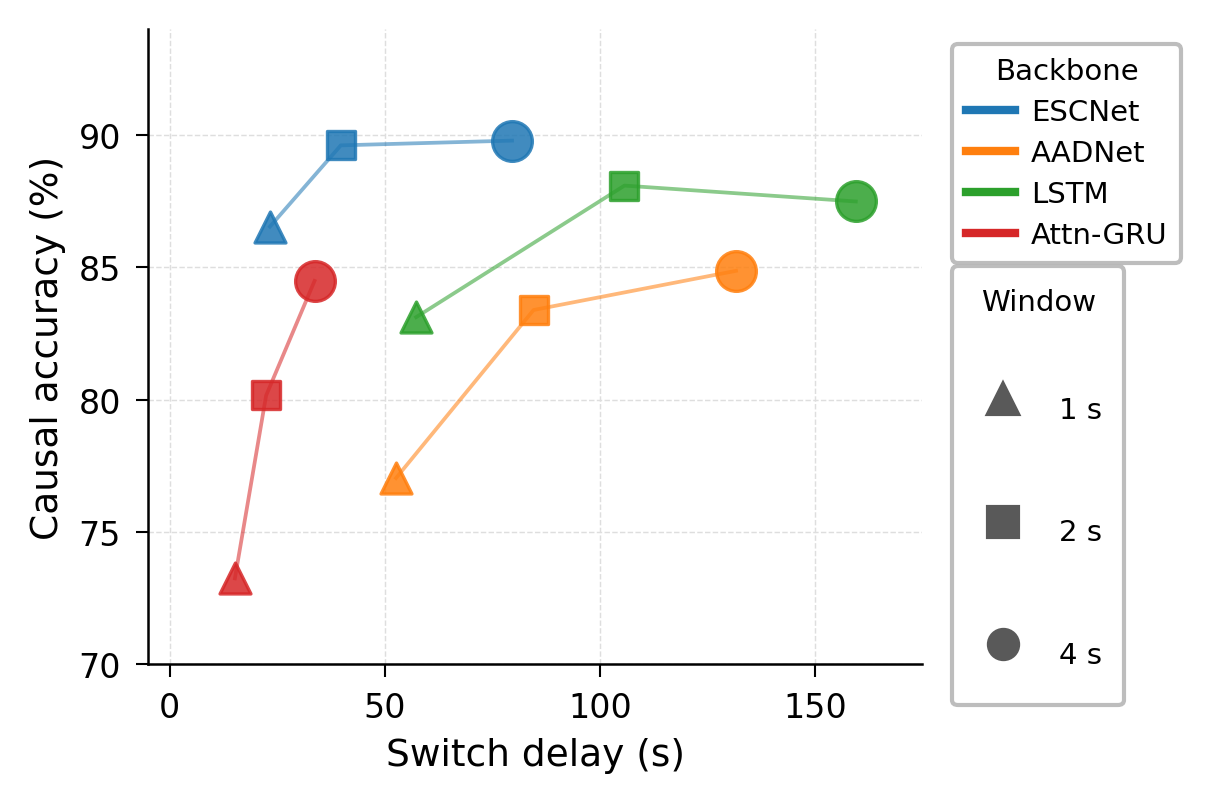}
\caption{AVGC causal-accuracy vs. switch-delay trade-off under CRF training.}
\label{fig:window-tradeoff}
\end{figure}

Figure~\ref{fig:window-tradeoff} makes this trade-off directly visible by plotting CRF causal accuracy against switch delay for every backbone at each window length, with bubble size encoding the window length and connecting lines tracing each backbone's $1$\,s$\to$2\,s$\to$4\,s trajectory. Longer windows generally provide stronger acoustic and neural evidence, but they also delay detection of genuine attention switches. The effect of sequence training is backbone- and window-dependent: at $1$\,s, CRF training improves both causal and non-causal decoding for all four backbones, whereas at longer windows some gains diminish or reverse for AADNet and LSTM. With ESCNet as the primary instantiation, CRF reaches $86.5\%$, $89.6\%$, and $89.8\%$ causal accuracy at $1$, $2$, and $4$\,s, respectively, while the mean switch delay increases from $23.3$ to $79.6$\,s. This configuration provides the strongest tested accuracy--delay balance, whereas Attn-GRU remains preferable when minimum delay is the overriding requirement. These results show both the applicability of CRF training across heterogeneous emissions and the dependence of its practical gain on emission quality.

\subsection{Static Evaluation on KUL and USTC}
Evaluations on the KUL and USTC datasets test whether Markov training helps beyond explicit switching scenarios. In these static AAD benchmarks, the attended speaker is stable within a trial under the standard protocol, so improvements cannot be explained by switch-specific behavior alone. Gains instead indicate that CRF training learns emissions better suited to persistent state trajectories. Table~\ref{tab:static-main} reports both datasets at the $1$\,s decision window, where weak per-window evidence provides a stringent test of sequence training; all models are evaluated using the same protocol and participant-level details are omitted for space.

\begin{table}[t]
\caption{Static AAD results on the KUL and USTC datasets at the $1$\,s decision window. All accuracies (\%) are participant-level means over LOTO folds. Post-Causal uses fixed $q_0=10^{-3}$; CRF-Causal learns $q$ from the same initialization. Both use causal (online) forward filtering and $q_{\tau}=q\tau$.  \textbf{Bold}: best CRF-Causal on each dataset.}
\label{tab:static-main}
\centering
\footnotesize
\begin{tabular}{llcccc}
\toprule
\multicolumn{2}{c}{Model} & Raw$\uparrow$ & Post-Causal$\uparrow$ & CRF-Causal$\uparrow$ & Gain (\%) \\
\midrule
\multirow{4}{*}{\rotatebox{90}{KUL}} &ESCNet & 54.9 & 82.0 & \textbf{87.7} & $+$5.6 \\
&LSTM      & 54.4 & 76.9 & 82.1 & $+$5.2 \\
&Attn-GRU  & 51.1 & 55.1 & 55.3 & $+$0.2 \\
&AADNet    & 52.5 & 74.8 & 79.7 & $+$4.9 \\
\midrule
\midrule
\multirow{4}{*}{\rotatebox{90}{USTC}} &ESCNet & 56.4 & 80.9 & \textbf{82.9} & $+$2.0 \\
&LSTM      & 54.8 & 70.9 & 76.3 & $+$5.4 \\
&Attn-GRU  & 52.9 & 57.5 & 58.0 & $+$0.5 \\
&AADNet    & 53.2 & 69.8 & 74.4 & $+$4.6 \\
\bottomrule
\end{tabular}
\end{table}

Table~\ref{tab:static-main} shows that CRF-Causal decoding improves over Post-Causal decoding for every backbone on both KUL and USTC. The two paths use the same window-scaled Markov structure, while the conventional HMM keeps the base rate fixed and CRF learns it jointly with the emissions. The strongest absolute results are obtained with ESCNet, which reaches $87.7\%$ on KUL and $82.9\%$ on USTC, with gains of $+5.6\%$ and $+2.0\%$ over its post-hoc HMM baselines. LSTM and AADNet also benefit consistently, whereas Attn-GRU changes by only $0.2\%\sim 0.5\%$. Together with the fixed-rate AVGC control, these results indicate that sequence-level training is useful beyond explicit switching tasks, although its practical value depends on the emission backbone.

\begin{figure*}[!h]
\centering
\begin{subfigure}{\linewidth}
\centering
\includegraphics[width=\linewidth]{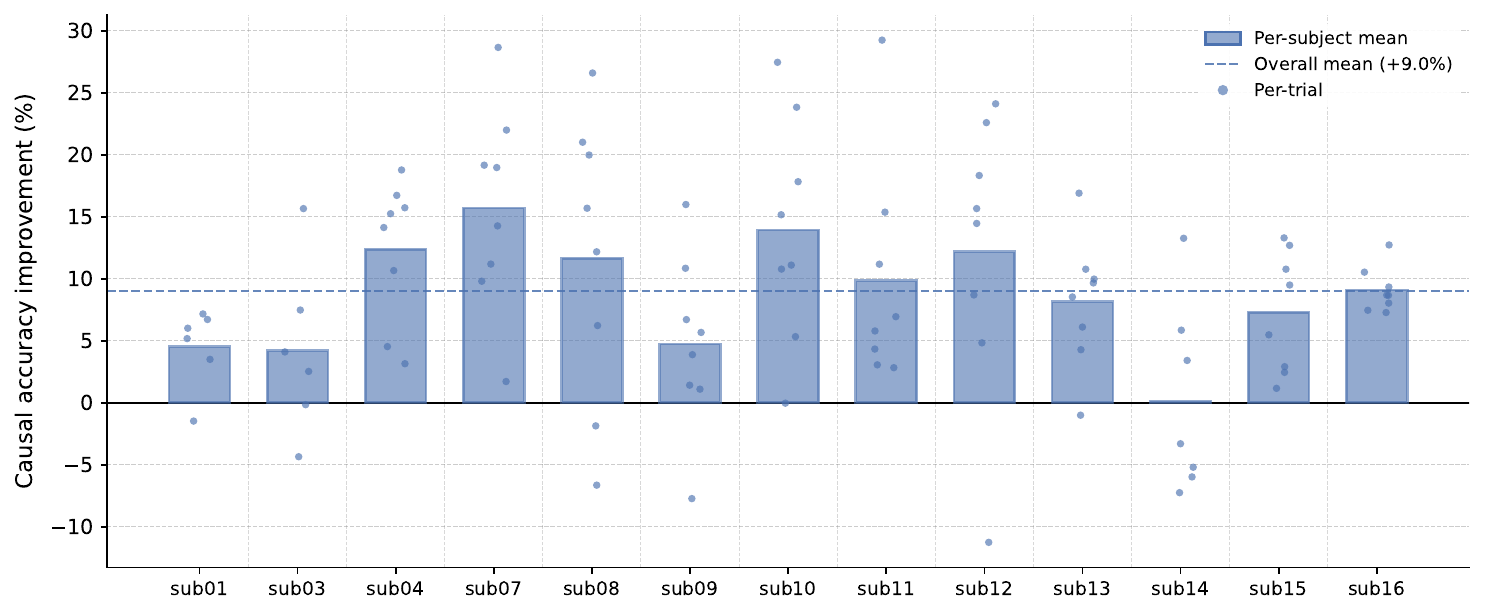}
\caption{Causal accuracy improvement.}
\label{fig:paired-subjects-causal}
\end{subfigure}
\par\bigskip
\begin{subfigure}{\linewidth}
\centering
\includegraphics[width=\linewidth]{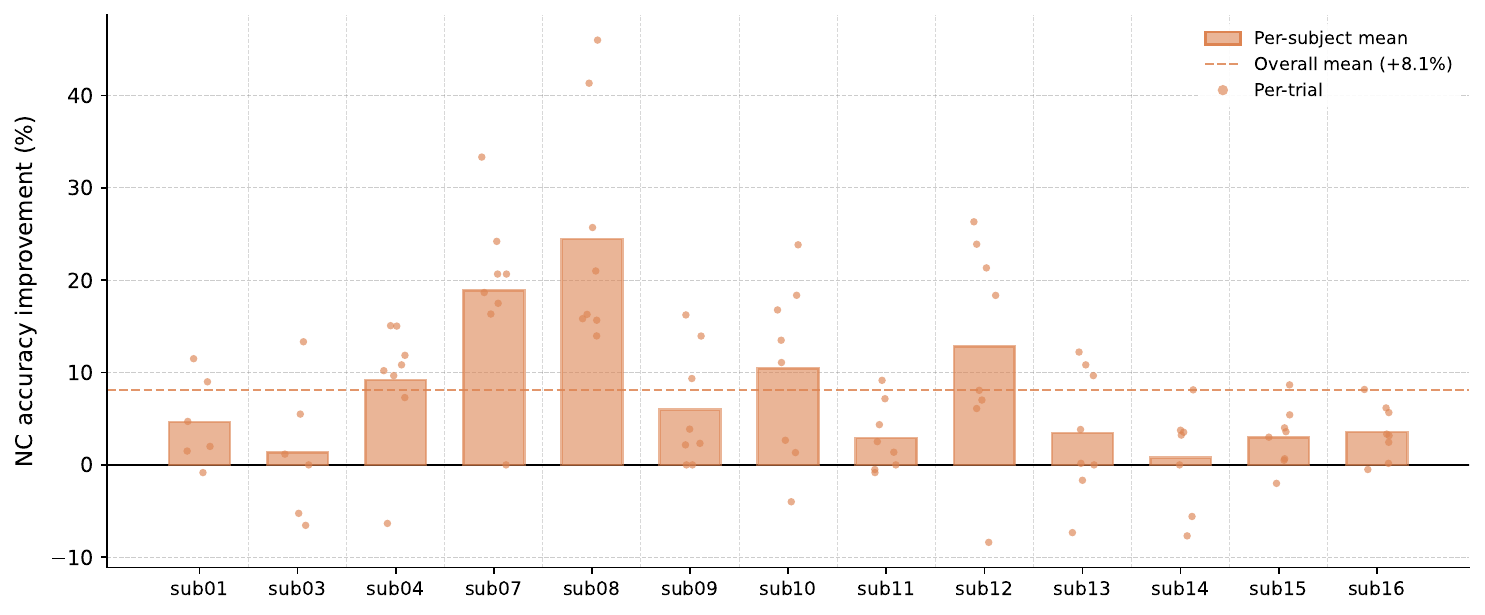}
\caption{Non-causal accuracy improvement.}
\label{fig:paired-subjects-fb}
\end{subfigure}
\caption{Per-participant Acc. improvement of CRF over HMM-post on AVGC using ESCNet emissions ($1$\,s window). Bar, dot and dashed line show participant-mean, fold-level and overall-mean gains (\%), respectively.}
\label{fig:paired-subjects}
\end{figure*}

Fig.~\ref{fig:paired-subjects} reports the paired participant-level improvement of the complete CRF framework over fixed-rate HMM-post on AVGC using ESCNet emissions with a $1$\,s window. Both methods use the same participants, folds, test windows, and backbone, while differing in sequence-aware backbone training and transition-rate learning. Subplots (a) and (b) show causal and non-causal improvement, respectively; bars are participant means over held-out folds, dots are fold-level improvements, and the dashed line is the overall mean. The fixed-rate comparison in Table~\ref{tab:ablations} separately controls for transition-rate learning.
All $13$ participants show a positive mean improvement in both modes: the causal mean is $+9.0\%$ (from $0.1\%$ to $15.7\%$), and the non-causal mean is $+8.1\%$ (from $0.8\%$ to $24.5\%$). Thus, the aggregate gain is not driven by only a few participants. The effect is nevertheless heterogeneous: participant sub14 has a small and variable causal gain, with fold-level differences from $-7.3\%$ to $13.3\%$, and several individual folds have negative differences, including one sub12 fold at $-11.3\%$. End-to-end training is thus consistent at the participant-average level but does not improve every held-out fold.

\subsection{Ablation and Analysis}

Table~\ref{tab:ablations} separates the contributions of temporal smoothing, sequence-level emission learning, and transition-rate adaptation. The CE-only model, which classifies each window independently, achieves $55.7\%$ accuracy. Applying a fixed-rate post-hoc HMM to the same independently trained emissions raises causal accuracy to $77.5\%$. When the transition rate is held at the same value, $q=q_0=10^{-3}$, end-to-end CRF training further raises causal accuracy to $86.0\%$ and non-causal accuracy from $84.3\%$ to $92.2\%$. These improvements of $8.5$ and $7.9$ percentage points, respectively, cannot be attributed to a different transition matrix and therefore isolate the benefit of using sequence-level supervision to learn the emissions. Because the decision window is $1$\,s, the corresponding per-window switching probability is also $q_{\tau}=10^{-3}$.

Adapting the transition rate alone has little effect on decoding accuracy. Starting from $q_0$, CRF training learns a mean base rate of $7.6\times10^{-4}$. When this learned rate is applied to the independently trained CE emissions, causal accuracy changes only from $77.5\%$ to $78.0\%$, while the mean switch delay changes from $9.3$ to $10.4$\,s. Under non-causal decoding, accuracy remains $84.3\%$, and the mean switch delay decreases from $14.4$ to $12.8$\,s. Thus, changing the transition rate without retraining the emissions does not reproduce the accuracy gain of the end-to-end CRF.

Jointly learning the transition rate within the CRF likewise provides only a small change beyond fixed-rate CRF training. It increases causal accuracy from $86.0\%$ to $86.5\%$ and non-causal accuracy from $92.2\%$ to $92.4\%$; the corresponding mean switch delays change from $21.8$ to $23.3$\,s and from $17.3$ to $17.4$\,s, respectively. Removing CE warm-up still outperforms the post-hoc HMM in accuracy, but increases the mean switch delay to $26.5$\,s for causal decoding and $29.8$\,s for non-causal decoding. This behavior suggests that sequence supervision without local initialization produces overly persistent predictions. Overall, the ablation attributes the main accuracy gain to sequence-aware emission learning rather than to transition-rate adaptation, while CE warm-up improves responsiveness to genuine attention switches.

\begin{table}[t]
\caption{Analysis of CRF training on AVGC using ESCNet and $1$\,s windows. NC denotes non-causal decoding using the complete trial. Here, $q$ is the base switching rate and equals the per-window probability because $\tau=1$\,s. The learned-rate post-HMM applies the mean rate learned by CRF, $7.6\times10^{-4}$, to independently trained CE emissions. Accuracy is the participant-level mean over full trials, and switch delay is the all-trial mean from Eq.~\eqref{eq:switch-delay}, including undetected switches.}
\label{tab:ablations}
\centering
\footnotesize
\begin{tabular}{lcccc}
\toprule
\multirow{2}{*}{Ablation}& \multicolumn{2}{c}{Causal} & \multicolumn{2}{c}{NC} \\
\cmidrule(lr){2-3}\cmidrule(lr){4-5}
 & \shortstack{Acc.(\%)} & \shortstack{Delay(s)} & \shortstack{Acc.(\%)} & \shortstack{Delay(s)} \\
\midrule
CE (raw) & 55.7 & -- & -- & --   \\
Post-HMM (fixed $q=q_0$) & 77.5 & 9.3 & 84.3 & 14.4 \\
Post-HMM (learned $q$) & 78.0 & 10.4 & 84.3 & 12.8 \\
CRF (no warm-up) & 82.3 & 26.5 & 89.3 & 29.8  \\
CE + CRF (fixed $q=q_0$) & 86.0 & 21.8 & 92.2 & 17.3 \\
CE + CRF (learned $q$) & 86.5 & 23.3 & 92.4 & 17.4  \\
\bottomrule
\end{tabular}
\end{table}
\subsection{Statistical Testing}
\label{sec:stat-test}

Finally, we use paired statistical tests because every method is evaluated on the same participants, folds, and test windows. The participant is therefore the unit of analysis, and CRF and HMM-post are paired rather than independent conditions. In these primary comparisons, HMM-post uses fixed $q_0$, whereas CRF learns $q$ from the same initialization; Table~\ref{tab:ablations} provides the complementary fixed-rate comparison. Confirmatory testing uses ESCNet as the primary instantiation of the framework; the remaining backbones assess transfer of the training principle and are compared descriptively. This focused hypothesis family avoids pooling significance tests across four architecturally different backbones.

For each dataset--window combination, we average accuracy over held-out folds within each participant and compute the paired difference between CRF and HMM-post. We report the mean difference, a $95\%$ confidence interval based on the $t$-distribution, Cohen's $d_z$ for paired samples, and a two-sided Wilcoxon signed-rank $p$-value. The hypothesis family contains eight tests: three AVGC window lengths under two inference modes and one causal comparison for each of KUL and USTC. We apply the Holm--Bonferroni procedure across these eight tests to control the family-wise type I error rate.

\begin{figure*}[t]
\centering
\includegraphics[width=\linewidth]{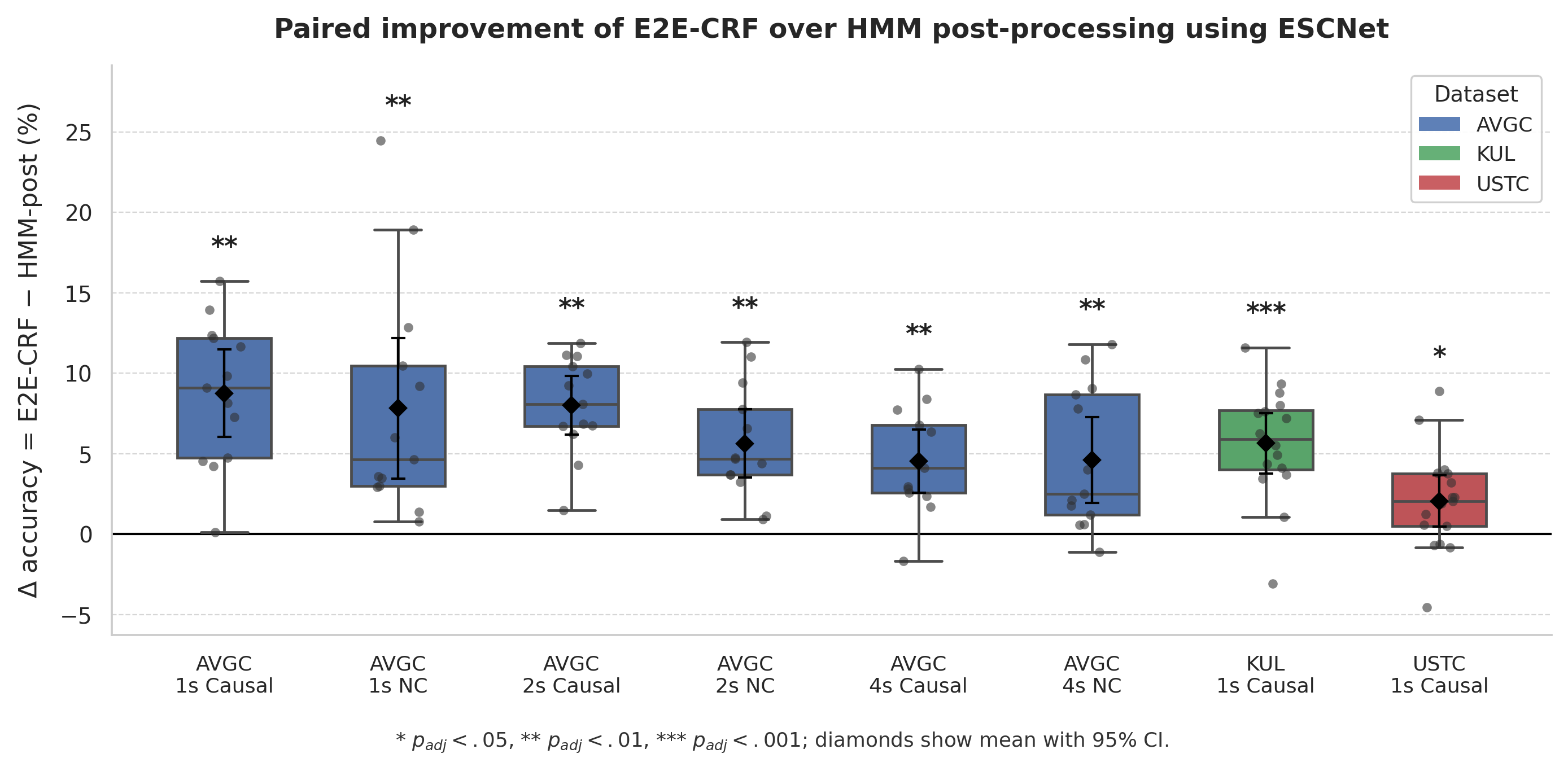}
\caption{Paired statistical test of CRF versus HMM-post. Positive values indicate that end-to-end Markov training improves over HMM post-processing, and vice versa. Significance stars use Holm-Bonferroni corrected $p$-values across all eight tests.}
\label{fig:stat-test}
\end{figure*}
Fig.~\ref{fig:stat-test} visualizes the paired participant-level differences of these tests. Each box summarizes $\Delta$ accuracy (CRF minus HMM-post) for one primary-backbone condition, points show individual participants, diamonds show the mean with a $95\%$ confidence interval, and stars denote Holm--Bonferroni-corrected significance. Using the primary instantiation, the paired analysis shows that the gains from CRF training are statistically reliable across the dynamic AVGC benchmark and both static datasets. Across AVGC, CRF improves causal and non-causal decoding at each window length (about $+4.4\%$ to $+9.0\%$ from the rounded means in Table~\ref{tab:avgc-main}, all corrected $p<0.01$), showing that sequence-level training remains effective when attention switches must be tracked. The same direction holds in the static setting: CRF improves causal decoding by $+5.6\%$ on KUL (corrected $p<0.001$) and by $+2.0\%$ on USTC (corrected $p=0.017$). The smaller USTC gain suggests that post-hoc HMM smoothing is already a strong baseline there, but the positive paired effect indicates that end-to-end Markov training still provides a measurable improvement.

\section{Conclusion}

This work reframes AAD as attention-state sequence estimation rather than independent short-window classification. We proposed ESCNet to extract time-preserving EEG--speech correlation evidence and directly form two-state logits from the difference between candidate correlation scores. We further proposed an end-to-end CRF framework that treats such neural logits as emissions of a two-state attention model and trains them with a sequence objective under the same type of temporal prior used by HMM post-processing. Together, these components couple interpretable local neural tracking evidence with trajectory-level supervision instead of using a Markov model only as a post-hoc smoother. On both dynamic and static datasets, the obtained results support the usefulness of learning emissions under a sequence objective. With ESCNet as the primary emission backbone, CRF improves over fixed-rate HMM post-processing across AVGC window lengths and inference modes, reaching $86.5\%$ causal and $92.4\%$ non-causal accuracy with $1$\,s windows while quantifying the accompanying accuracy--delay trade-off near attention switches. On KUL and USTC, the same instantiation can improve causal decoding over the corresponding fixed-rate post-hoc HMM baselines, and paired tests show that these gains are statistically reliable. The proposed CRF training algorithm is not tied to the proposed emission architecture, which is also applicable to other existing AAD models, although the gains are backbone-dependent. These support both ESCNet as an effective correlation-based instantiation and CRF as a general strategy for aligning local emissions with the temporal structure of auditory attention.

This study might be bounded by its evaluation setting, as AVGC is the only dataset with annotated within-trial attention switches, whereas KUL and USTC evaluate persistence under static trial labels rather than naturalistic multi-switch listening. Even the best causal AVGC configuration has a mean switch delay of $23.3$\,s, which may need to be further optimized for real-time HAs. Moreover, the evaluated participants are normal-hearing, and the experiments assume access to two candidate speech streams and controlled scalp-EEG preprocessing. The reported algorithmic gains do not yet establish improved speech perception or rehabilitation outcomes for hearing-impaired users. Clinical translation will require lower-latency decoding and validation with hearing-impaired participants, wearable or ear-centered EEG, natural acoustic scenes, variable numbers of talkers and online adaptation.
\bibliographystyle{IEEEtran}
\bibliography{citations}


\end{document}